# Nanodosimetry – on the tracks of biological radiation effectiveness

**Hans Rabus**

Physikalisch-Technische Bundesanstalt, 38116 Braunschweig, Germany

It is well known that the biological effectiveness of a certain absorbed dose of ionizing radiation depends on the radiation quality, i. e. the spectrum of ionizing particles and their energy distribution [1], [2], [3]. As has been shown in several studies, the biological effectiveness is related to the pattern of energy deposits on the microscopic scale, the so-called track structure [4]. Clusters of lesions in the DNA molecule within site sizes of few nanometers play a particular role in this context [4], [5], [6].

A first approach to measure track structure of ionizing radiation with nanometric resolution was proposed already in 1975 [7]. However, the extension of microdosimetric measurements to site sizes of few nanometer dimension was facing the fundamental problem that in such small sites the number of interactions is too low, such that the assumption fails that the imparted energy is the number of ionizations multiplied by a simple conversion factor [8]. Therefore, the development of methods for measuring track structure details with nanometric resolution required a change of paradigm, namely restricting the characterization of track structure to its ionization component [9], [10].

In should be noted in this context that the term nanodosimetry is used in the literature in different meanings. These include ongoing endeavors to extend microdosimetry into the nanometer range to below 100 nm site sizes [11], [12] as well as simulation studies of various kinds that are not focused on quantities that are directly measurable. In this paper we use the term nanodosimetry for studies of charged particle track structure, considering the stochastics of ionizations in nanometric targets.

The quantity of interest is the relative frequency distribution of the so-called ionization cluster size, i.e. the number of ionizations inside a considered target (often called the 'site'). As is illustrated in Figure 1, the ionization cluster size distribution (ICSD) varies with the geometrical position of the target with respect to the particle track. Furthermore, it also depends on the size and composition of the target and, most importantly, on the radiation quality. The ICSD can also be characterized by its statistical moments or the complementary cumulative frequencies of ionization clusters exceeding a certain minimum size [13].

Around the turn of the century, three different types of nanodosimeters have been developed to measure the frequency distribution of ionization cluster size [14]. All devices are gas counters that simulate nanometric sensitive volumes based on a density scaling principle [15]. They

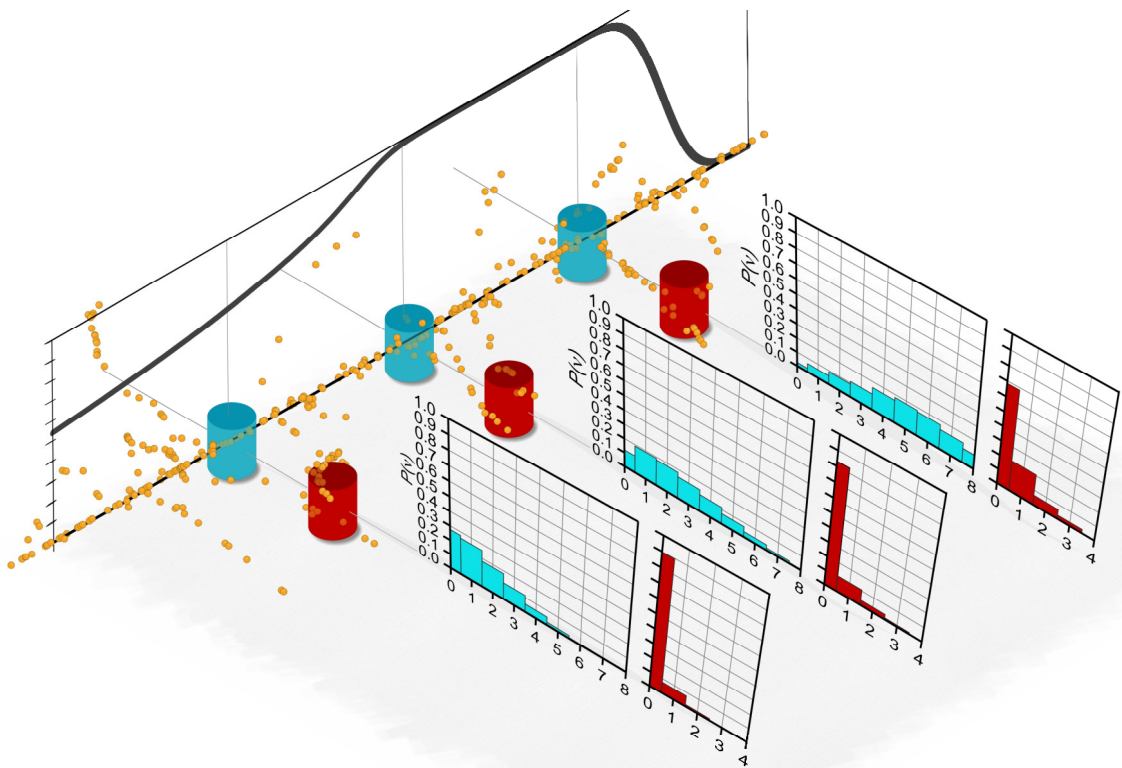

Figure 1. Schematic illustration of nanodosimetric ionization cluster size (ICS) distributions in a spread-out Bragg peak (SOBP) of protons in water. The orange spheres indicate the loci of ionizing interactions of the proton or of the emitted secondary electrons. The blue and red cylinders represent nanometric targets located in the core and in the penumbra region of the track, respectively. The histograms with the blue and red columns show the corresponding relative frequency distributions of ICS (up to ICS of 8 and 4, respectively) as obtained by Monte Carlo simulation with the Geant4-DNA code for target cylinders of 2.3 nm diameter and 3.4 nm height. As can be seen, the ICS distributions shift towards larger cluster values across the SOBP, where this variation is more pronounced in the track core region, indicating a potential rise in differ in the details of their operation principles and in the size of the (single) simulated nanometric site [14].

Within the European project BioQuaRT [16], [17], an intercomparison of the three nanodosimeter types was conducted for a number of ion types and energies. Despite the large differences between the ionization cluster-size distributions measured with the three devices for each radiation quality (ion type and energy), the parameters of the measured distributions were found to follow a universal relation (see Fig. 2 in reference [18]). Furthermore, this universal relation was also found to reproduce for a number of cell lines the dependence on linear energy transfer of the effective cross sections for inducing the biological endpoint of cell inactivation after ion irradiation (Figs. 4 to 7 in [18]).

In this context, the relevant nanodosimetric parameter depends on the repair capacity of the cells and the value of absorbed dose, and there is basically only one scaling factor between cumulative frequencies and the cross sections which turns out to be close to the cell nucleus cross sectional area. For details the reader is referred to reference [18].

In a subsequent work by Conte et al., these finding were exploited even further to establish a link between measured nanodosimetric parameters of ion track structure and the coefficients of the linear-quadratic model for cell survival [19]. This progress has stimulated investigations into the potential use of nanodosimetric parameters in treatment planning, as these offer the advantage of measurable quantities linked to the biological outcome of the irradiation [20], [21], [22].

Despite the undeniable and impressive evidence found by Conte et al. [18], [19], a major concern that is often raised is the implication that radiation interaction with a single nanometric target should decide the fate of the irradiated cell. It is still an open issue to resolve how these finding can be reconciled with the overwhelming evidence of the importance of indirect radiation effects (i.e. via radiation chemistry) and existing evidence of several relevant scales for radiation effectiveness [23].

In fact, also the BioQuaRT project developed a multi-scale model that included microdosimetry and nanodosimetry and the possibility of different relevant target sizes depending on the biological endpoint considered [17]. At least two similar approaches have been developed independently by other groups [24], [25]. Furthermore, theoretical investigations suggest that biological radiation effects originate in the interaction of different DNA lesions [26].

In consequence, respective new developments have been started in the field of experimental nanodosimetry that aim at measuring the correlations of cluster-size distributions induced within a particle track in two separate nanometric targets in proximity [27], [28] or at obtaining a 3D image of the nanometric particle track structure for track segments of few 100 nm in length [29], [30], [31]. The latter would to some extent bridge towards microdosimetric measurements at the few-hundred nanometer regime [12].

A recent simulation study by Selva et al. investigated the exploitation of the universal nanodosimetric relation [18] in radiotherapy by using (still only hypothetical) solid-state nanodosimeters that contain multiple targets of nanometric site size [32]. In fact, a first attempt at developing such nanodosimetric detectors with true nanometric dimensions has been recently made by using electrical nano-circuits with DNA-based nanowires [33]. Both approaches imply to some extent a need for revision of the concepts of nanodosimetry [32]. However, this would actually be aligned with the presently most advanced approach to using nanodosimetry for predicting relative biological effectiveness (RBE) in radiotherapy treatment planning, the so-called track-event theory of Besserer and Schneider [34], [35].

Although there have also been first approaches applying nanodosimetry with brachytherapy photon sources [36], the first field where nanodosimetry will be used in treatment planning seems to be radiotherapy using protons or ion beams. For the latter, the variation of the RBE is already accounted for in treatment planning [37], while for protons a constant RBE-weighting factor of 1.1 is used and the likely RBE variation is accounted for by considering the so-called biological range uncertainty [38], [39].

Several major challenges still have to be mastered before nanodosimetric treatment planning will come into routine clinical practice. One is the development of nanodosimeters that are suited for daily quality assurance measurements on a routine basis as opposed to the quite bulky and sophisticated instruments used for nanodosimetric measurements so far. A second one relates to simulations and to the development of methods for fast computation of relevant nanodosimetric parameters along the entire particle track [40]. Establishing a 'gold standard' Monte Carlo code for track structure simulations (including the chemical and biochemical reactions ensuing the physical radiation action) and the experimental validation of its entrance data as well as its intermediate results also seems a necessity. Eventually, the assessment of the uncertainties of nanodosimetric quantities obtained from measurements [41] and simulations [42] is of paramount importance.

**Acknowledgements**

I thank S. A. Ngcezu for providing the data for the figure and Heike Nittmann for help with preparing the illustration.

Corresponding Author: Hans Rabus, Email: hans.rabus@ptb.de